# ARTICLE    OPEN

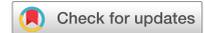

# The neuroscience of advanced scientific concepts

Robert A. Mason [1,2]✉, Reinhard A. Schumacher [2] and Marcel Adam Just [1,2]

Cognitive neuroscience methods can identify the fMRI-measured neural representation of familiar individual concepts, such as apple, and decompose them into meaningful neural and semantic components. This approach was applied here to determine the neural representations and underlying dimensions of representation of far more abstract physics concepts related to matter and energy, such as fermion and dark matter, in the brains of 10 Carnegie Mellon physics faculty members who thought about the main properties of each of the concepts. One novel dimension coded the measurability vs. immeasurability of a concept. Another novel dimension of representation evoked particularly by post-classical concepts was associated with four types of cognitive processes, each linked to particular brain regions: (1) Reasoning about intangibles, taking into account their separation from direct experience and observability; (2) Assessing consilience with other, firmer knowledge; (3) Causal reasoning about relations that are not apparent or observable; and (4) Knowledge management of a large knowledge organization consisting of a multi-level structure of other concepts. Two other underlying dimensions, previously found in physics students, periodicity, and mathematical formulation, were also present in this faculty sample. The data were analyzed using factor analysis of stably responding voxels, a Gaussian-naïve Bayes machine-learning classification of the activation patterns associated with each concept, and a regression model that predicted activation patterns associated with each concept based on independent ratings of the dimensions of the concepts. The findings indicate that the human brain systematically organizes novel scientific concepts in terms of new dimensions of neural representation.

*npj Science of Learning* (2021)6:29 ; https://doi.org/10.1038/s41539-021-00107-6

## INTRODUCTION

Physics is the fundamental science that studies the interactions of matter and energy across all scales of space and time. From antiquity to the present, science has struggled to find the best possible conceptual framework for understanding the physical world. While basic physics concepts such as *velocity* and *torque* have perceptual counterparts that can provide a neural basis for their representation, it is unclear how the brain has accommodated to represent non-intuitive or counter-intuitive concepts involving the subatomic, quantum, and cosmological realms. Here we characterize the representation of the most advanced scientific concepts in the field of physics, as they occur in the brains of university faculty physicists. Recent functional magnetic resonance imaging of brain function has enabled the study of how various types of mundane, everyday concepts are neurally and cognitively represented in the human brain. We now apply this approach to understanding the underlying neural and semantic organization of highly abstract contemporary scientific concepts in the brains of active physicists.

The brain organization of meaning refers to the underlying dimensions of representation that are used in the human brain. For example, physical objects such as hand tools are neurally represented in terms of how one's body interacts with the object, what the object looks and feels like, and what is purpose. Each of these underlying dimensions of meaning correspond to the activation pattern in a particular brain subsystem, such as the wielding of a hammer being represented in the brain's motor representation of arm movement. In the domain of classical physics, the multiple underlying dimensions of concepts such as *gravity* and *wavelength* and *light* include a dimension of energy flow, a dimension of periodicity, and a dimension of visualized causal motion[1]. But now we ask, what are the dimensions of representation of contemporary scientific concepts such as *dark matter* or *multiverse*, concepts without any associated perceptual or motor information?

This study investigated the cognitive and neural dimensions that underlie the representation of post-classical concepts like *dark matter*, which differ from classical concepts. The "classical" domain of physics includes fruitful concepts dating from antiquity, the middle ages, the renaissance, and up to about the end of the 19th century. Classical concepts are associated with discoveries and insights by thinkers such as Aristarchus, Archimedes, Alhazen, Newton, Faraday, and Maxwell (among many others). Classical physics can be said to capture the operations of nature using matter, energy, forces, and fields that are straightforward to visualize and to grasp intuitively with some training. For instance, while a force cannot directly be seen, physicists can imagine "an arrow of force" of definite size and direction acting upon a material object. Similarly, the energy contained in a system cannot be seen in a pictorial sense, but the intuitive idea of a conserved quantity that flows from one specific form to another can be related to human perceptual experience. Most fMRI research on physics has focused on intuitive physics[2] and classical Newtonian forces[3–5].

Near the beginning of the twentieth century, however, a paradigm shift arose requiring a radical conceptual change. This shift occurred with the introduction of successful ideas that were not amenable to direct visualization or ordinary intuition. These included quantization of various classical concepts, the relativization of space and time, the plethora of sub-atomic particles that obey rules for which there are no classical analogs, unexpected emergent phenomena with no classical analogs, and new celestial building blocks for which there are no classical analogs. These

[1]Center for Cognitive Brain Imaging, Carnegie Mellon University, Pittsburgh, PA 15213, USA. [2]Department of Psychology, Carnegie Mellon University, Pittsburgh, PA 15213, USA. ✉email: rmason@andrew.cmu.edu

| **Table 1.** The 45 stimulus concepts used in the study, drawn from various physics sub-fields. |
| --- |
| *Post-classical concepts* |
| boson, fermion, muon, neutrino, particle decay; |
| anti-particle, cosmology, dark matter, multiverse, quasar; |
| gamma ray, inertial frame, Lorentz invariant, simultaneity, tachyon; |
| coherence, commutator, duality, wave function |
| *Classical concepts* |
| acceleration, centripetal force, gravity, torque, velocity; |
| direct current, electric field, force, potential energy, voltage; |
| frequency, light, radio waves, sound waves, wavelength; |
| buoyancy, Coriolis force, fractal dimension, Lagrangian, Hamiltonian; |
| precession, canonical ensemble, conduction, crystal lattice, diamagnetism, insulator |
| Concepts were either defined as either classical (Newtonian or Maxwellian) or post-Classical. The concepts are grouped here by physics subfield. The partitioning of concepts in this study into classical and post-classical is somewhat arbitrary (grouping label designated by italics): it was generated by one of the authors (RS), a particle physics experimentalist with over 30 years of experience teaching physics to undergraduate and graduate students. |

concepts arose not from perceptual experience, but from the generative capabilities of the human brain.

Such "post-classical" concepts as *dark matter* or *fermion* are distinguished for our purpose as being (a) less easy to visualize directly as mental images, (b) possibly failing to have an everyday intuitive aspect that "makes sense", even after repeated exposure, and (c) require the introduction of new rules of behavior ("laws") that have no correspondence to ordinary life experience. Such concepts can be thought of as more abstract, in the sense that they are less related to perception, but simply labeling them as abstract fails to specify the cognitive content of the abstraction. Our study begins to provide some of that specification.

The present study obtained the fMRI-based activation patterns for each of 45 concepts (shown in Table 1), and then applied factor analysis to the activation levels of representative voxels to find the main factors or dimensions along which the activation patterns are organized. (We henceforth use dimensions and factors interchangeably, preferring factors when referring to the factor analysis outcomes and dimensions when referring to neural organization.) One subset of the concepts was drawn from "classical" concepts familiar to physics undergraduates, while another subset was drawn from concepts that are generally post-classical, familiar to Ph.D. physicists irrespective of their research specialties, and often requiring ideas from quantum physics or being of a speculative nature. As a caution, we note that the concepts labeled as post-classical often also have classical components too. For example, a neutrino has both a classical aspect, being a particle like any other, flying through space with mass and momentum, but at the same time it has oscillatory quantum aspects that are unexplainable classically.

Using this approach of factor analyzing the neural signatures of physics concepts, we demonstrate that there are four describable factors underlying the neural representations of both classical and post-classical physics concepts, constituting a kind of orthogonal vector basis of the neural representations. Furthermore, we show that (a) each physics concept has a distinct associated activation pattern that can be accurately identified by a statistical classifier, (b) that the activation patterns for each concept are measurably common across scientists, (c) that the activation pattern for a concept that has been excluded from the modeling of the other 44 concepts can be accurately predicted by a model that uses expert behavioral ratings of the concept with respect to the postulated underlying dimensions, and (d) that the faculty's representations of basic physics concepts can be reliably distinguished from those of students. These findings provide a first characterization of the neural organization for representing advanced physics concepts.

## RESULTS

Five types of results are reported below: 1. Description of the neural dimensions underlying advanced physics concepts in physicists; 2. Testing of the dimension descriptions using a generative model to predict the brain activation of "new" individual concepts that have been held out from the modeling; 3. Identification of individual concepts from their neural signatures; 4. Assessment of the commonality of the neural signatures across physicists; and 5. Machine learning to differentiate faculty from students based on their neural representations of classical physics concepts.

## Neural dimensions of physics concept knowledge in experts

Physics concepts are represented in terms of a consistent and identifiable set of neural dimensions in the brains of experts, namely Carnegie Mellon Physics Department faculty. Four semantic factors emerged from a factor analysis procedure (see Methods section) of the consistently activated voxels. These factors constitute an orthogonal set of dimensions that collectively underpinned the neural representations of all of the concepts.

The interpretations of the dimensions are:

1. relating to a measurable magnitude;
2. relating to a mathematical formulation;
3. entailing a repeated systematic change over time as in periodicity or wave-related concepts;
4. relating to classical physics as opposed to post-classical intangible but consilient ideas.

The interpretations of the dimensions were initially developed by one of the authors (RS), based primarily on how the concepts were ordered by their factor scores along each dimension, with particular attention paid to the concepts near the extremes of the dimension. This was followed by a rating of the concepts along each of the four dimensions (as interpreted) by six Carnegie Mellon physics faculty members who had not participated in the fMRI study. They rated how strongly each of the 45 concepts was related to each of the four hypothesized dimensions using a 7-point rating scale. There were two quantitative measures of how well the hypothesized dimensions accounted for the fMRI activation data, as described below: one measure was the correlation between the expert ratings of the concepts relative to each factor and the concepts' factor scores on that factor, and the other measure was the accuracy of prediction of the activation pattern of held-out concepts based on their ratings. The dimensions are described in more detail below.

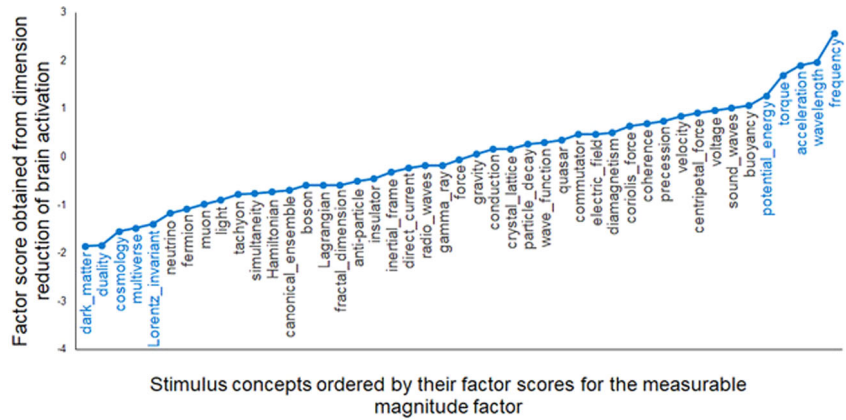

**Fig. 1 The fMRI-based factor scores for the 45 concepts on the measureable magnitude dimension (vertical axis).** The contrast between the concepts at the two ends of the dimension (highlighted in blue text) indicates the nature of the dimension.

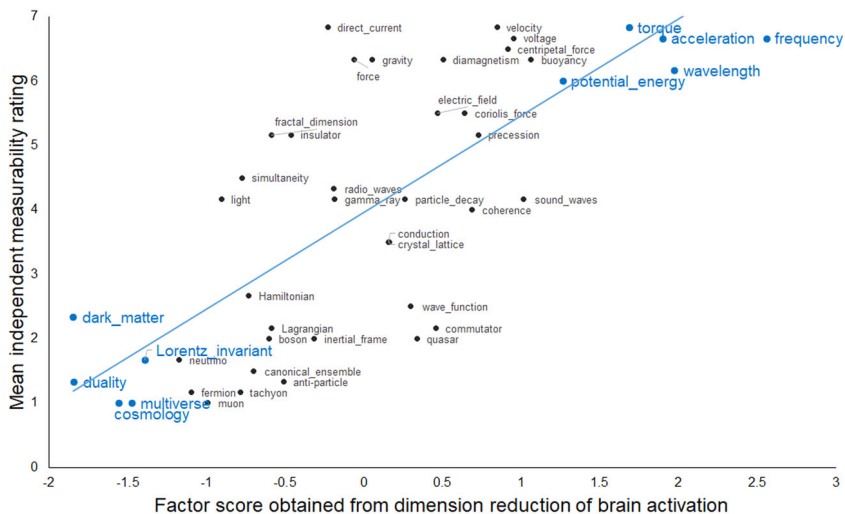

**Fig. 2 Factor scores for the measurable magnitude factor from the fMRI brain data (x-axis) were well correlated ($r = 0.74$) with the mean expert ratings of the concepts with respect to that factor (y-axis).** The concepts at the extreme ends of the factor score distribution that are influential for factor interpretation are highlighted in blue in a larger font.

*Measurable magnitude.* Concepts with high factor scores on this dimension include *frequency, wavelength, acceleration*, and *torque*. In contrast, low factor scores were associated with non-numerical concepts like *dark matter, duality, cosmology*, and *multiverse*. The factor scores on the measurable magnitude dimension for all of the concepts are shown in Fig. 1. The concepts at the extreme ends of the dimension clearly reflect the interpretation, as they did for the other three dimensions.

To assess how well the interpretation of the dimension fits the data, correlations between the concepts' factor scores and the ratings provided by the independent set of physics faculty were computed for each factor. The correlation (Pearson's *r*) between the concepts' measurable magnitude ratings and their factors scores was 0.74 for all 45 concepts and 0.97 for the 10 concepts at the extremes of the dimension, as shown in Fig. 2 (both of these correlation coefficients are significant at $p < 0.01$). The ratings and the factor scores are particularly well correlated for the concepts at the two ends of the distribution; the relationship remains but is less strong for the concepts in the middle parts of the dimension. The brain locations associated with this factor were mostly left-lateralized, and more lateralized than for the other factors. The clusters of voxels that loaded highly on this measurable magnitude factor were located primarily in left middle to superior temporal gyrus, left inferior parietal, left superior parietal, and left intraparietal sulcus (IPS), and to a lesser degree, left hemisphere precuneus, left superior frontal gyrus, and occipital regions. The centroids of the clusters associated with each of the factors are shown in Supplementary Table 1.

*Mathematical formulation.* Concepts with high factor scores on this dimension, including *commutator, Lagrangian*, and *Hamiltonian*, were strongly associated with mathematical expressions, equations, or transformations. Concepts with low factor scores on this dimension included *sound waves, quasar*, and *direct current*, concepts that do not require a mathematical formulation to be grasped. The correlation between the concept ratings along this dimension and the factors scores was 0.45 for all 45 concepts and 0.87 for the 10 concepts at the dimension extremes (both were significant at $p < 0.01$). The clusters of voxels with high loadings on this factor were located in left superior parietal, left supplementary motor, right precentral sulcus and right pars triangularis.

*Periodicity/wave-related.* The concepts that had high factors scores on this dimension included *wave function, light, radio waves, gamma ray*, and *coherence*. The concepts with low factor scores included *inertial force, buoyancy*, and canonical ensemble, none of which are periodic. The correlation between the ratings and the factors scores was 0.47 for all 45 concepts and 0.94 for the 10 concepts at the dimension extremes (both were significant at $p < 0.01$). The clusters of voxels with high factor loadings on this factor were located in right inferior frontal gyrus (including a pars triangularis and a pars opercularis cluster), right middle temporal, and right angular gyrus/inferior parietal regions, with additional clusters located in left inferior temporal and bilateral precuneus.

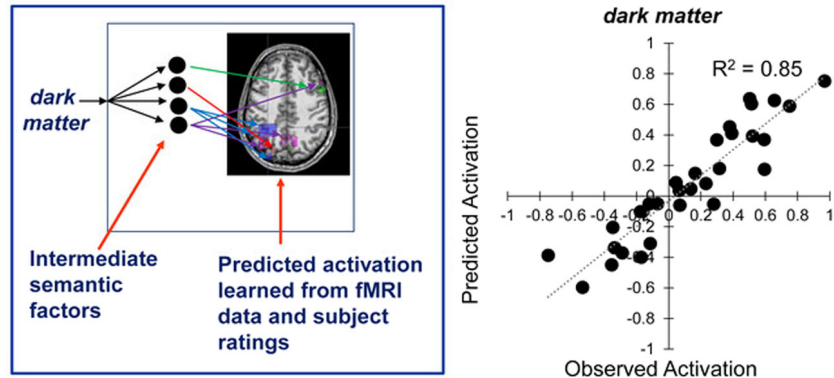

**Fig. 3 The predictive model presented graphically and as data.** The left panel is a schematic representation of the predictive model. The right panel shows a scatterplot of observed and predicted activation values in the 30 factor-related clusters for a sample concept, *dark matter*, where $R^2 = 0.85$. For this illustration, the predictive model was applied to a mean dataset obtained by averaging the activation data of all participants, and developing the mapping from the ratings of the other 44 concepts along the four main factors to the mean activation level of the 30 cluster locations associated with the factors. The resulting regression weights were then applied to the ratings for the left-out (45th) concept (dark matter) to predict its activation values in those 30 locations.

*Classical vs. post-classical.* This factor was interpreted as a categorical dimension that distinguished between classical and post-classical concepts. The concepts with high factor scores (classical concepts) included *velocity, acceleration, force, potential energy*, and *torque*, but also other classical concepts such as *canonical ensemble, light*, and *direct current*, as well as associated mathematical tools like *Lagrangian* and *Hamiltonian*. The concepts with low factor scores included post-classical concepts such as *anti-particle, multiverse*, and *tachyon*. The correlation between the ratings and the factor scores was 0.35 for all 45 concepts and 0.95 for the extreme 10 concepts at the dimension extremes (these two correlation coefficients were significant at $p < 0.02$ and $p < 0.01$, respectively). In the discussion section, the cognitive processes involved in thinking about post-classical concepts are categorized into four classes (reasoning about intangibles, assessing consilience, causal reasoning, and knowledge management) that each correspond to specific brain locations (voxel clusters).

*Word length.* The factor scores of the concepts along this dimension were highly correlated with the character-length of the concept label ($r = 0.81$; $p < 0.001$). The voxels with high factor loadings on this factor were located almost exclusively in the occipital lobe where first-order visual word-perception processing occurs. This factor simply reflects the neural encoding of the visual form of the word that names the concept. This finding provides a validity check on the method of relating factor scores to properties of the concepts.

*Unlabeled factor.* One additional dimension accounted for above-threshold variance in the first-level factor analysis and re-emerged in the second level analysis (see "Methods" section). This dimension was also present when rotations other than varimax (e.g., equimax, parsimax, and quartrimax) were investigated. However, physics experts were unable to develop an interpretation that accounted for the concepts' factor scores. Concepts associated with one extreme of this factor were *anti-particle, precession, Lagrangian, acceleration, commutator*, and *fractal dimension*. Voxels with high factor loadings on this dimension were located in the periphery of the visual cortex (occipital lobe, cuneus, and left occipital parietal junction) and in a right middle frontal region, typically active in working memory tasks.

**Predicting the activation pattern of held-out individual concepts, based on the concept ratings along the four main dimensions**

A regression model developed a mapping from the mean ratings (averaged across raters) of 44 of the 45 concepts along the four main dimensions to the mean activation level of 30 cluster locations associated with the four factors. The mapping was based on the activation data from 7 of the 10 participants. (The 30 cluster locations were the minimal set obtained from a factor analysis of the data from the three participants with the highest concept classification accuracies. The predictive modeling was performed on the remaining 7 participants' data. The weights from this regression model, obtained from the mapping between concept ratings and activation levels in the 30 locations for 44 of the concepts, were used to predict the activation pattern of the left-out (45th) concept in the 30 factor-related locations, as shown schematically on the left side of Fig. 3.

The predictive model was evaluated in two ways: (1) the similarity of the model predictions to the observed activation patterns, which was assessed using $R^2$ (the goodness of fit as the proportion of the variation in the observed activation data explained by the predictions of the model); and (2) the ability to distinguish among concepts, which was assessed using classification accuracy based on the distance between the predicted and observed activation of each concept. On the first measure, the model had a good fit to the data as indicated by a mean $R^2$ of 0.82 (averaged over the 45 predictions and seven test participants, with a standard deviation of 0.11 across all participants and concepts; the mean $R^2$ was 0.81 when the 3 participants whose data established the factor locations were included). The mean observed and predicted activation values in the 30 clusters are shown on the right side of Fig. 3 for a sample concept, *dark matter*.

The second measure of the predictive model, classification accuracy, indicated the predictive model's ability to distinguish among the concepts. A classifier was trained to identify each concept, based on the proximity (vector correlation distance) between the observed activation for a test concept and the set of predicted activation patterns for all of the concepts. (Here, and throughout the paper, normalized rank accuracy of the classification performance was computed as the normalized rank of the correct label in the classifier's posterior-probability-ordered list of

    

all the classes. If the classification were operating at chance level, one would expect a mean normalized rank accuracy of 0.50, indicating that the correct word appeared on average between the 22nd and 23rd position in the classifier's output of a ranked list of the 45 items). The mean normalized rank accuracy across the seven participants was 0.70 ($p < 0.001$ by permutation test), and 0.69 when all ten participants were included. Thus the predictive model based on the semantic factors makes accurate and reliably discriminative predictions.

### Within-participant identification of physics concepts from their neural signatures in a bottom-up model

In addition to the classification above, based on the predictive model, a more conventional discriminative classifier (Gaussian Naïve Bayes) was trained to identify a concept from its activation pattern. The classifier was trained on a subset (4 of the 6 presentations) of the fMRI data from each participant and then tested on an independent subset (the mean of the remaining two presentations) of the same participant's data, using as features the 120 most stable voxels in each fold (iteration) of the cross-validation protocol. The 45 physics concepts were classified with a mean normalized rank accuracy of 0.78 (range = 0.57–0.92). (When occipital cortex voxels were excluded, to limit the impact of the non-semantic word length factor, the mean classification accuracy decreased to 0.75. Recall that the predictive model, which also excluded voxels related to visual perception of the concept label achieved a mean rank accuracy of 0.70.) Individual participant rank accuracies were reliably above a $p < 0.001$ chance level for 8 of the 10 participants and above a $p < 0.05$ level for the remaining two participants. The three participants whose data were used to establish factor locations for the predictive modeling had accuracies of 0.87, 0.88, and 0.92. Thus the concepts are reliably identifiable from their neural signatures in this bottom-up analysis.

### Commonality of the neural signatures across physicists

The commonality of the neural representations across participants was assessed by training a classifier on data (voxel activation levels) for each of the concepts from all but one participant and then testing it on the data of the left-out participant. The features were the activation levels at 120 voxel locations that had a consistent profile across the nine participants in the training set (i.e. voxels with high pairwise correlations between participants' mean activation levels over presentations). The mean cross-participant classification rank accuracy to identify the 45 concepts for the 10 participants was 0.70 (range = 0.53–0.81). These reliable results (nine participants had accuracies greater than chance at $p < 0.01$ and one participant at $p < 0.05$) indicate the commonality across participants of the neural representations of these physics concepts.

### Differentiating faculty from students based on their neural representations of classical physics concepts

In all, 15 of the 45 concepts that had been presented to the faculty group were basic classical concepts that had also been presented to a student group in a previous study[1]. These concepts were *acceleration, centripetal force, gravity, torque, velocity, direct current, electric field, force, potential energy, voltage, frequency, light, radio waves, sound waves,* and *wavelength*. A machine-learning classifier was trained to identify whether the neural representations of these classical concepts had come from a faculty member or a student. The brain locations (voxel clusters) that were used as features in this classification were obtained by taking the union of the factor locations obtained in two separate factor analyses performed on the data of each group on the voxel activation patterns of the 15 elementary concepts.

When the classifier was trained on the group membership data for all 15 concepts from all but one participant, it achieved a mean accuracy of 0.79, correctly classifying the group membership for 10 of 14 faculty and 5 of 9 students. (The four members of the student group misclassified as faculty were all recent graduates who were enrolled in graduate programs. Their conceptualization of the 15 basic concepts were apparently more similar to the faculty than to the undergraduate students. The five correctly classified students were all undergraduates.) The finding that the neural representations of the more advanced students were more similar to the faculty than were the representations of more junior students suggests that the neural representations change systematically with additional learning or with academic progress. Thus it may be possible to assess the degree of a student's learning in terms of some property of the neural representations and more generally to investigate the relationship of individual differences in academic achievement to properties of neural representations, as others have suggested[4,6].

To identify the set of concepts whose activation best discriminated between the groups, a reiterative stepwise classification procedure (analogous to stepwise regression) was performed[7], yielding four most discriminating concepts: *sound waves, radio waves, gravity,* and *force.* Some of the discriminating concepts had high factor scores on the periodicity factor, which was associated with similar locations in the two groups (bilateral inferior temporal and right parietal). However, the clusters at these locations were smaller and more spatially compact in the faculty group, possibly due to increased neural efficiency[8]. The factor locations that were specific to the faculty group included right frontal and right temporo-parietal junction clusters, possibly the result of activation of remote associations[9]. In summary, faculty physicists' neural representations of some basic concepts were reliably distinguishable from students' representations.

### DISCUSSION

This study identified the content of the neural representations in the minds of physicists considering some of the classical and post-classical physics concepts that characterize their understanding of the universe. In this discussion, we focus on the representations of post-classical concepts, which are the most recent and most abstract and have not been previously characterized psychologically. The neural representations of both the post-classical and classical concepts were underpinned by four underlying neurosemantic dimensions, such that these two types of concepts were located at opposite ends of the dimensions. The neural representations of classical concepts tended to be underpinned by underlying dimensions of measurability of magnitude, association with a mathematical formulation, having a concrete, non-speculative basis, and in some cases, periodicity. By contrast, the post-classical concepts were located at the other ends of these dimensions, stated initially here in terms of what they are not (e.g. they are not periodic and not concrete). Below we discuss what they are.

The main new finding is the underlying neural dimension of representation pertaining to the concepts' presence (in the case of the classical concepts) or absence (in the case of post-classical concepts) of a concrete, non-speculative basis. The semantic characterization of this new dimension is supported by two sources of converging evidence. First, the brain imaging measurement of each concept's location on this underlying dimension (i.e. the concepts' factor scores) converged with the behavioral ratings of the concepts' degree of association with this dimension (as we have interpreted it) by an independent group of physicists. (This type of convergence occurred for the other three dimensions as well.) Second, the two types of concepts have very distinguishable neural signatures: a classifier can very accurately distinguish the mean of the post-classical concepts' signatures from the mean of the classical concepts' within each participant, with a grand mean accuracy of 0.93, $p < 0.001$.

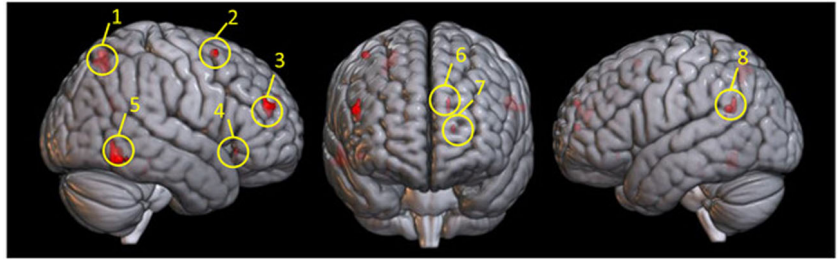

**Fig. 4   Factors locations associated with the post-classical end of the classical vs. post-classical dimension (i.e. those whose activation was increased for post-classical concepts).** The factor clusters are encircled and numbered for ease of reference in the text and their centroids are included in Table 2. These locations correspond to the four classes of processes evoked by the post-classical concepts.

### Neural representations of post-classical concepts
As physicists ventured into conceptually new territory in the 20th century and developed new post-classical concepts, their brains organized the new concepts with respect to a new dimension that had not played a role in the representation of classical concepts.

To describe what mental processes might characterize the post-classical end of this new dimension, it is useful to consider what attributes of the post-classical concepts could have led to their being neurally organized as they are and what cognitive and neural processes might operate on these attributes. Previously mentioned was that post-classical concepts often involve their immeasurability and their lower likelihood of being strongly associated with a mathematical formulation and periodicity, both of which are attributes that are often absent from post-classical concepts.

### A new neurosemantic organization for post-classical concepts
More informative than the "absent" attributes are four types of cognitive processes evoked by the post-classical concepts: (1) Reasoning about intangibles, taking into account their separation from direct experience and their lack of direct observability; (2) Assessing consilience with other, firmer knowledge; (3) Causal reasoning about relations that are not apparent or observable; and (4) Knowledge management of a large knowledge organization consisting of a multi-level structure of other concepts.

In addition to enabling the decoding of the content of the participants' thoughts, whether they were thinking of *dark matter* or *tachyon* for example, the brain activation patterns are also informative about the concomitant psychological processes that operate on the concepts, in particular, the four processes listed above are postulated to co-occur specifically with the post-classical concepts. The occurrence of these processes was inferred from these locations of the voxel clusters associated with (having high loadings on) the classical/post-classical factor, specifically the factor locations where the activation levels increased for the post-classical concepts. (These voxel clusters are shown in Fig. 4, and their centroids are included in Table 2). Inferring a psychological process based on previous studies that observed activation in that location is called reverse inference. This can be an uncertain inferential method because many different processes or tasks can evoke activation at the same location. What distinguishes the current study are several sources of independent converging evidence, in conjunction with the brain locations associated with a factor (and not simply observed activation), indicating a particular process.

First, a statistically reliable decoding model predicted the activation levels for each concept in the factor locations, based on independent ratings of the concepts with respect to the postulated dimension/factor. The activation levels of the voxels

**Table 2.** Brain regions associated with the post-classical concepts, grouped by hypothesized cognitive process (designated in italic).

| | Cluster number on figure | $x$ | $y$ | $z$ | # of Voxels |
|---|---|---|---|---|---|
| *Reasoning about intangibles* | | | | | |
| Left Supramarginal | 8 | −56 | −51 | 23 | 11 |
| Right inferior temporal | 5 | 59 | −59 | −11 | 15 |
| *Assessing consilience* | | | | | |
| Left medial frontal | 6 | −12 | 54 | 26 | 4 |
| Right middle frontal | 2 | 41 | 8 | 58 | 3 |
| *Causal reasoning* | | | | | |
| Left supramarginal | 8 | −56 | −51 | 23 | 11 |
| Right superior parietal | 1 | 26 | −65 | 52 | 15 |
| Right middle frontal | 3 | 47 | 43 | 22 | 7 |
| Right inferior orbital frontal | 4 | 45 | 28 | −35 | 2 |
| *Knowledge management* | | | | | |
| Left medial frontal | 6 | −12 | 54 | 26 | 4 |
| Left medial frontal | 7 | −15 | 53 | 10 | 2 |
| *Unassigned clusters* | | | | | |
| Right fusiform | Not shown | 28 | −35 | −17 | 2 |

The activation levels of the voxel clusters in these regions increased for the post-classical concepts with extreme factor scores associated with this dimension. Since regions are multipotent, some are grouped under more than one process.

in the factor locations were systematically modulated by the stimulus set, with the post-classical concepts, a specific subset of the stimuli eliciting the highest activation levels in these locations, resulting in the highest factor scores for this factor. Thus these brain locations were associated with an activation-modulating factor, not with a stimulus or a task. Second, the processes are consistent with the properties participants reported to have associated with the post-classical concepts. These properties provide converging evidence for these four types of processes occurring. For example, the concept of *multiverse* evoked properties related to assessing consilience, such as "a hypothetical way to explain away... constants." Another example is that *tachyons* and *quasars* were attributed with properties related to reasoning about intangibles, such as "quasi-stellar objects". Third, the processes attributed to the factor locations were based not simply on an occasional previous finding, but on the large-scale meta-analysis (the Neurosynth database, Yarkoni et al.[10]) using the "association based test" feature. The association between the location and the process was based on the cluster centroid locations; particularly relevant citations are included in the factor descriptions. Each of the four processes is described in more detail below.

*Reasoning about intangible concepts.* The nature of many of the post-classical concepts entails the consideration of alternative possible worlds. The post-classical factor location in the right temporal area (shown in cluster 5 in Fig. 4) has been associated with hypothetical or speculative reasoning in previous studies. In a hypothetical reasoning task, the left supramarginal factor location (shown in cluster 8) was activated during the generation of novel labels for abstract objects[11]. Additionally, the right temporal factor location (shown in cluster 5) was activated during the assessment of confidence in probabilistic judgments[12].

*Assessing consilience.* Another facet of post-classical concepts is that they require the unknown or non-observable to be brought into consilience with what is already known. The right middle frontal cluster (shown in cluster 2) has been shown to be part of a network for integrating evidence that disconfirms a belief[13]. This consilience process resembles the comprehension of an unfolding narrative, where a new segment of the narrative must be brought into coherence with the parts that preceded it. When readers of a narrative judge the coherence of a new segment of text, the dorsomedial prefrontal cortex location (shown in cluster 6) is activated[14]. This location is associated with a post-classical factor location, as shown in Fig. 4. Thus understanding the coherence of an unfolding narrative text might involve some of the same psychological and neural consilience-seeking processes as thinking about concepts like *multiverse*.

*Causal reasoning.* Thinking about many of the post-classical concepts requires the generation of novel types of causal inferences to link two events. In particular, the inherent role of the temporal relations in specifying causality between events is especially complex with respect to post-classical concepts. The temporal ordering itself of events is frame-dependent in some situations, despite causality being absolutely preserved, leading to counter-intuitive (though not counter-factual) conclusions. For example, in relativity theory the concept of *simultaneity* entails two spatially separated events that may occur at the same time for a particular observer but which may not be simultaneous for a second observer, and, even the temporal ordering of the events may not be fixed for the second observer. Because the temporal order of events is not absolute, causal reasoning in post-classical terms must eschew inferencing on this basis, but must instead rely on new rules ("laws") that lead to consilience with observations that indeed can be directly perceived.

Another example, this one from quantum physics, concerns a particle such as an electron that may be conceived to pass through a small aperture at some speed. Its subsequent momentum becomes indeterminate in such a way that the arrival location of the particle at a distant detector can only be described in probabilistic terms, according to new rules ("laws") that are very definite but not intuitive. The perfectly calculable non-local "wave function" of the particle-like object is said to "collapse" upon arrival in the standard Copenhagen interpretation of quantum physics. Increasingly elaborate probing of physical systems with one or several particles, interacting alone or in groups with their environment, has for decades elucidated and validated the non-intuitive new rules about limits and about causality in the quantum world. The fact that new rules regarding causal reasoning are needed in such situations was described as "the heart of quantum mechanics" and as containing "the only mystery" by Richard Feynman[15].

Generating causal inferences to interconnect a sequence of events in a narrative text evokes activation in a right temporal and right frontal location (shown in clusters 3 and 4) which are post-classical factor locations[16–18] as shown in Fig. 4. Causal reasoning accompanying perceptual events also activates a right middle frontal location (shown in cluster 3) and a right superior parietal location (shown in cluster 1)[19]. Notably, the right parietal activation is the homolog of a left parietal cluster associated with causal visualization[1] found in undergraduates' physics conceptualizations, suggesting that post-classical concepts may recruit right hemisphere homologs of regions evoked by classical concepts. Additionally, a factor location in the left supramarginal gyrus (shown in cluster 8) is activated in causal assessment tasks such as determining whether the causality of a social event was person-based (being a hard worker) or situation based (danger)[20].

*Knowledge management.* Although we have treated post-classical concepts such as multiverse as a single concept, it is far more complex than *velocity*. *Multiverse* entails the consideration of the uncertainty of its existence, the consilience of its probability of existence with measurements of matter in the universe, and the consideration of scientific evidence relevant to a multiverse. Thinking about large, multi-concept units of knowledge, such as the schema for executing a complex multi-step procedure evokes activation in medial frontal regions (shown in cluster 6)[21,22]. Reading and comprehending the description of such procedures (read, think about, answer questions, listen to, etc.) requires the reader to cognitively organize diverse types of information in a common knowledge structure. Readers who were trained to self-explain expository biological texts activated an anterior prefrontal cortex region (shown in cluster 7 in Fig. 4) during the construction of text models and strategic processing of internal representations[23].

This underlying cognitive function of knowledge management associated with the post-classical dimension may generate and utilize a structure to manage a complex array of varied information that is essential to the concept. This type of function has been referred to as a Managerial Knowledge Unit[22]. As applied to a post-classical concept such as a *tachyon*, this knowledge management function would contain links to information to evaluate the possibility of the existence of tachyons, hypothetical particles that would travel faster than light-speed in vacuum. The concept invokes a structured network of simpler concepts (*mass, velocity, light*, etc.) that compose it. This constitutes a knowledge unit larger than a single concept.

## All four dimensions underlie the neural representations

Although the discussion has so far focused on the most novel dimension (the classical vs. post-classical), all four dimensions together compose the neural representation of each concept, which indicates where on each dimension a given concept is located (assessed by the concept's factor scores). The bar graphs of Fig. 5 show how the concepts at the extremes of the dimensions can appear at either extreme on several dimensions. These four dimensions are:

1. the classical vs. post-classical dimension, as described above, which is characterized by contrasting the intangible but consilient nature of post-classical concepts versus the quantifiable, visualizable, otherwise observable nature of classical concepts.
2. the measurability of a magnitude associated with a concept, that is, the degree to which it has some well-defined extent in space, time, or material properties versus the absence of this property.
3. the periodicity or oscillation which describes how many systems behave over time versus the absence of periodicity as an important element.
4. the degree to which a concept is associated with a mathematical formulation that formalizes the rules and

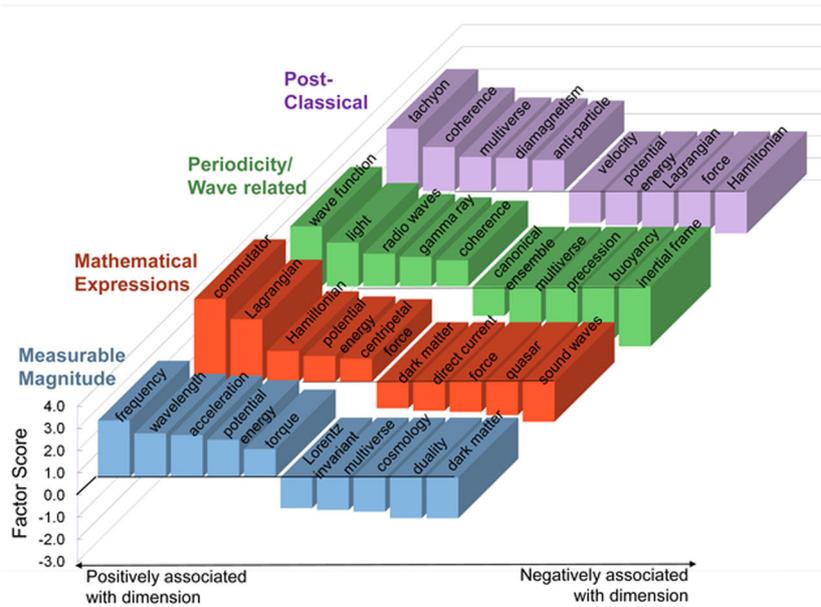

**Fig. 5  Factor scores of concepts at the extremes of the four factors.** A concept may have a high factor score for more than one factor; for example, *potential energy* appears as measurable, mathematical, and on the classical end of the post-classical dimension. In contrast, *multiverse* appears as non-measurable, non-periodic, and post-classical.

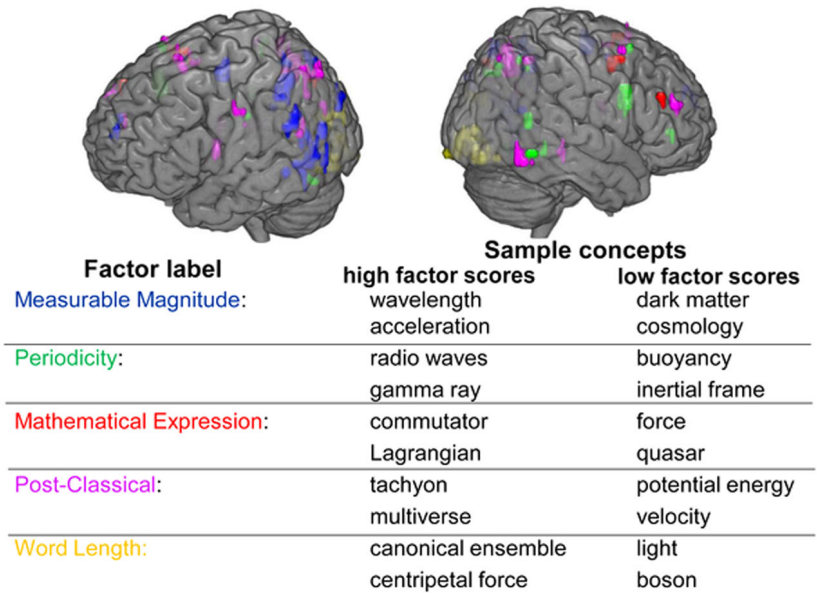

**Fig. 6  Factor locations for the five factors (35 voxel clusters) are depicted on a rendered brain.** Colors differentiate the factors and greater color transparency indicates greater depth. Sample concepts from the two ends of the dimensions are listed. The post-classical factor locations include those whose activations were high for post-classical concepts (their locations are shown in Fig. 4) as well as those locations whose activations were high for classical concepts.

principles of the behavior of matter and energy versus being less specified by such formalizations.

The locations of the clusters of voxels with high loadings on each of the factors are shown in Fig. 6.

Classical concepts with high factor scores on the measurability factor, such as *frequency, wavelength, acceleration,* and *torque,* are all concepts that are often measured, using devices such as oscilloscopes and torque wrenches, whereas post-classical concepts such as *duality* and *dark matter* have an uncertainty of boundedness and no defined magnitude resulting in factor scores at the other end of the dimension. This factor is associated with parietal and precuneus clusters that are often found to be activated when people have to assess or compare magnitudes of various types of objects or numbers[24–26], a superior frontal cluster that exhibits higher activation when people are comparing the magnitudes of fractions as opposed to decimals[27], and an occipital-parietal cluster (dorsolateral extrastriate V3A) that activates when estimating the arrival time of a moving object[28]. Additional brain locations associated with this factor include left supramarginal and inferior parietal regions that are activated during the processing of numerical magnitudes;[26] and left intraparietal sulcus and superior parietal regions activated during the processing of spatial information[29]. This factor was not observed in a previous study that included only classical concepts and hence the factor would not have differentiated among the concepts[1].

The mathematical formulation factor is salient for concepts that are clearly associated with a mathematical formalization. The three concepts that are most strongly associated with this factor, *commutator*, *Lagrangian*, and *Hamiltonian*, are mathematical functions or operators. Cluster locations that are associated with this factor include: parietal regions that tend to activate in tasks involving mathematical representations[30,31] and right frontal regions related to difficult mental calculations[32,33]. The parietal regions associated with the factor, which extend into the precuneus, activate in arithmetic tasks[34]. While most if not all physics concepts entail some degree of mathematical formulation, post-classical concepts such as *quasar*, while being measurable, are typically not associated with an algebraic formulation.

The periodicity factor is salient for many of the classical concepts, particularly those related to waves: *wave function, light, radio waves,* and *gamma rays*. This factor is associated with right hemisphere clusters and a left inferior frontal cluster, locations that resemble those of a similarly described factor in a neurosemantic analysis of physics concepts in college students[1]. This factor was also associated with a right hemisphere cluster in the inferior frontal gyrus and bilateral precuneus.

### Predicting the activation pattern of novel concepts

For all four underlying semantic dimensions, the brain activation-based orderings of the physics concepts with respect to their dimensions were correlated with the ratings of those concepts along those dimensions by independent physics faculty. This correlation makes it possible for a linear regression model to predict the activation pattern that will be evoked by future concepts in physicists' brains. When a new physics concept becomes commonplace, (such as a new particle category, say, magnetic monopoliae), it should be possible to predict the brain activation that will be the neural signature of the magnetic monopole concept, based on how that concept is rated along the four underlying dimensions.

The neurosemantic conceptual space defined by the four underlying dimensions includes regions that are currently sparsely populated by existing concepts, but these regions may well be the site of some yet-to-be theorized concepts. It is also possible that as future concepts are developed, additional dimensions of neural representation may emerge, expanding the conceptual space that underpins the concepts in the current study.

### EXTENSIONS AND LIMITATIONS

This account of the neural representation of physics concepts may generalize to other scientific disciplines, such as the biological sciences, where the concepts and dimensions presumably are related to the properties of living things and their transformations. One of the first documented neurosemantic disturbances associated with brain damage concerned the selective loss of knowledge in laypersons about living versus nonliving things[35,36]. This neural living-nonliving dimension may turn out to be central in representing scientific concepts among biological scientists. However, neuroimaging studies of advanced scientific thinking remain uncommon, and the existing studies have not examined the representations of concepts. For example, two fMRI studies of skilled mathematicians focused on the processes used in solving complex math problems[37,38] and on the execution of calculations[39].

The findings concerning physics faculty may have implications for the training of future physicists. If the underlying dimensions of neural representation of concepts in a scientific discipline are known, then it may be advantageous to intentionally teach these concepts using the underlying neurosemantic dimensions to describe, explain, and define the concepts. For example, the oscillatory nature of light and sound energy concepts could be one of their properties that are explicitly taught together. Indeed, at various universities there are undergraduate courses with titles similar to "Waves and Vibrations".

One limitation of these findings is that the observed association of the concepts with the four underlying dimensions is based on only the first four seconds of thought. Our findings indicate that it would be quite likely for a physicist thinking of *mass* to think of a concomitant magnitude, but much less so in the case of *electron*. However, given more time, even a concept like electron, which has no magnitude in the naïve sense, can be subsequently thought of in terms of the magnitude of its charge, magnetic moment, or velocity. Thus the neural representations that this study assessed are the first-order representations rather than extended associations that may eventually be evoked.

### CONCLUSION

Scientific thought is a product of a specialized component of human culture, built on a structured but plastic neural infra-structure. The neural representations of advanced physics concepts exhibit the brain's organization of the concepts in terms of its own representational schemas. Scientific progress built on this neural organization continually enlarges the conceptual grasp of the discipline and constitutes some of the most notable intellectual accomplishments of the human mind and brain.

### MATERIALS AND METHODS

#### Participants

#### Experimental paradigm

The stimuli were 45 physics terms with five concepts from each of nine physics topic areas: (particle/nuclear physics, astrophysics, condensed matter, special relativity, classical mechanics, quantum mechanics, elementary classical mechanics, elementary energy/electricity, and elementary light/sound). These concepts were selected to be representative of the knowledge of any Ph.D.-level academic physicist, irrespective of research specialization. The 15 concepts from three elementary categories were a subset of those included in a previous investigation of physics concepts in college students (11).

The set of 45 concepts was presented six times (in six different random permutation orders of the 45 items). Each concept label was visually presented on a video screen for 4 s during which the participant thought about the properties of the concept, followed by a 6 s rest period, during which the participant fixated on a shrinking and gradually disappearing blue ellipse displayed in the center of the screen. There were seven additional longer presentations (17 sec) of a shrinking ellipse distributed across the session to provide a baseline measure of brain activity.

### Task

The participants were instructed to actively and re-iteratively think about the properties of the presented concept. To promote their consideration of a consistent set of properties or features across the six presentations of each concept, participants were asked to write down two or three properties of their choosing for each item prior to the scanning session (for example, the properties for the term *velocity* might be vector quantity, movement related, and directional). Although this task is relatively easy, it is fairly demanding and could invite inattention. A sufficiently high level of classification accuracy (as described below) is used to ensure that all participants paid attention and performed the task throughout the experiment.

### fMRI procedures

Functional images were acquired on a Siemens Verio (Erlangen, Germany) 3.0T scanner at the Scientific Imaging and Brain Research Center of Carnegie Mellon University using a gradient echo planar imaging (EPI) pulse sequence with TR = 1000 ms, TE = 25 ms and a 60° flip angle. Twenty 5-mm-thick AC-PC aligned slices were imaged with a gap of 1 mm between slices using a 32-channel head coil. The acquisition matrix was 64 × 64 with 3.125-mm × 3.125-mm × 5.0-mm in-plane resolution. Images were corrected for slice acquisition timing, motion, and linear trend, and were normalized to the Montreal Neurological Institute (MNI) template without changing voxel size (3.125 × 3.125 × 6 mm) using SPM8 (Wellcome Dept. of Cog. Neurology). The gray matter voxels were assigned to anatomical areas using Anatomical Automatic Labeling (AAL) masks [31].

The percent signal change (PSC) relative to the fixation condition was computed at each gray matter voxel for each stimulus presentation. The main input measure for the subsequent analyses consisted of the mean activation level averaged over the four 1-s brain images acquired within a four second window, offset five seconds from the stimulus onset (to account for the delay in hemodynamic response). This measure has previously resulted in higher classification accuracies than other conventional measures.

### Selecting voxels with stable activation patterns

A stable voxel was defined as one that responded similarly to the 45-item stimulus set each time the set was presented. This measure is of a voxel's stability and was computed as the mean pairwise correlation between its set of 45 activation levels across all pairs of the presentations that served as training input for a given classification model. High stability is thus an analytic for the replicability of the voxel's semantic tuning curve. The voxel selection is based on only the training data for the model in each cross validation fold and is then applied to the test data. Additional details are listed in the Supplementary Methods.

### Factor analysis methods

Factor analysis was applied to the activation data for three purposes: to reduce the dimensionality of the neural activity associated with the 45 different stimulus items to a modest number of components, to determine the ordering of the items with respect to each factor by their factor scores, and to determine the multiple brain locations of the stable voxels that were most associated with each factor. A two-level exploratory factor analysis (FA) procedure was applied. First, the individual level FA applied to the individual data of the three most accurately classified participants (in the bottom-up classification of the concepts), as has been done in a previous study (3, 11). The goal of these individual factor analyses was to obtain similar outcomes (factor structures) in multiple participants by focusing on the individual datasets with the greatest systematicity. Second, a group level FA was performed on the aligned and merged data of these three participants. A Matlab implementation of a principal FA algorithm including varimax rotation, equivalent to the SAS v. 9.2 (http://www.sas.com), procedure was used. This FA procedure is described in detail elsewhere[41,42] and in the Supplementary Methods.

At the individual level, four separate factor analyses were performed on each participant using activation data from four regions, namely frontal, parietal, temporal-fusiform, and occipital lobes bilaterally (as defined in the Automated Anatomical Labeling (AAL) atlas[40]). The input data was the matrix of intercorrelations among the activation profiles (the vector of activation levels) across the 45 concepts of the 120 most stable voxels in each of these regions. The rationale for performing a separate FA in each lobe rather than one FA for the entire cortex was to prevent any of the regions from dominating the set of input stable voxels, as the occipital regions might have done otherwise. The choice of the particular number of voxels per region was motivated by similar analyses in other datasets where a total of 120 to 150 voxels produced a consistent range of maximal classification accuracy. The individual level FA's provided a set of 10 factors per lobe, each characterized by their factor scores for the 45 concepts and the factor loadings of the input voxels.

The group level factor analysis based on the output of the individual-level analyses was then used to identify factors that were common across regions and participants. The search for commonality of factors across regions was motivated by the assumption that a factor would be underpinned by a large-scale cortical network with representation in multiple and disparate brain regions. The input to the group-level analysis consisted of the factor scores for the 45 items associated with the ten dominant individual-level factors in each lobe obtained from the three most accurately classified participants. Voxels were uniquely assigned to one of the factors by using their highest (absolute value) loading above a 0.4 threshold. For each factor, the associated voxels tended to cluster in 4 to 12 different locations in the brain. Voxel clustering was performed by finding at least two neighboring voxels associated with a given factor. A cluster of two voxels is not likely to be spurious because their combined volume of 117 mm³ is substantial and more importantly that they were each stable in all three participants. The factors were aligned across the three participants by the correlations of their factor scores for the 45 items.

The output of the group level FA was a set of factors, a corresponding set of factor scores for the concepts, and clusters of voxels with high loadings on the factors. More specifically, the FA produced three interpretable semantic factors (measurable magnitude, periodicity, and mathematical expression) and one factor corresponding to the visual property of word length. (A factor was considered interpretable if some interpretation fairly clearly applied to the five items at each of the two extremes of the dimension.)

A second attempt was made to find additional interpretable factors by focusing on stable voxels clusters that had not been associated with any interpretable factor. These voxels' factor scores from the individual-level analyses were subjected to a second exploratory group-level factor analysis, resulting in the emergence of an additional interpretable factor (post-classical) as well as one remaining uninterpretable factor. The resulting set of factors from both group-level analyses (the initial group level FA on all clusters and the secondary group level FA on the any remaining uninterpretable clusters) included the four physics-related semantic factors, one uninterpretable factor, and a word length factor. Only the four semantic factors were used in analyzing the data of the other seven participants. Supplementary Table 1 lists the centroids of the 35 clusters associated with the semantic factors in the group-level FA.

The factors/dimensions emerging from these three participants' data were also present to a very large degree in the fMRI data of the other participants, such that the four semantic factors were reproduced in a separate verification factor analysis on the full set of ten participants. When the word length factor was included, 47 out of 50 possible participant-cluster regions were present in the individual participants' FA's.

### Machine learning analyses

Gaussian Naïve Bayes (GNB) classifiers were used to identify the 45 physics concepts (for an overview of the GNB classifier cross-validation as applied to fMRI data see[41] and the Supplementary Methods). The classifiers were trained using the activation levels of stable voxels from only a subset of the data (the training set), and then tested on the remaining independent data (the test set) using a cross-validation procedure. For the within-participant classification, the training set on each fold consisted of the data for each item (i.e. the activation levels of the selected voxels) from four of the six presentations and the test set consisted of the mean of the data from the remaining two presentations. For cross-participant classification, the classifier was trained on the data from nine participants and tested on the 10th, left-out participant. In the latter analysis, each participant's data was averaged over the six presentations. Then the 120 voxels with the most similar activation profiles across the 45 concepts (assessed with correlation) across the nine participants in the training set were selected as features for the classifier.

### Ratings task

An independent set of six faculty who were not participants in the fMRI study rated the 45 concepts with respect to the four interpreted semantic/

physics dimensions using a 7-point scale. Participants rated items along one dimension at a time, with three of the dimensions (measurable magnitude, periodicity, and mathematical) presented in a random order, and the classical-postclassical dimension last. Instructions described the dimension as precisely as possible, using as examples items that were not part of the stimulus set. For example, the instructions for the measurable magnitude dimension stated:

"This attribute refers to the degree to which a concept has a measurable magnitude. The concepts at one end clearly have a measurable magnitude (and should get a rating of "7") whereas the concepts at the other end of the dimension clearly do not have a well-defined magnitude and should get a rating "1". An example of a concept with a measurable magnitude is *mass*, which might receive a rating of "7" for this attribute. An example of a concept without a measurable magnitude is *electron*, which might receive a rating of "1." If a concept has some intermediate degree of relationship to magnitude (such as *divergence*), please give it a rating around the middle of the scale."

## Predictive modeling

In this model, the postulated relation of each concept to each of the four dimensions was estimated by the mean ratings described above. A linear regression model with four predictor variables developed a mapping between the ratings along the four dimensions (factors) of all but one concept and the mean fMRI activation level in each of the 30 factor clusters/locations for that concept. (The mean activation for a cluster was defined as the mean activation level of the five most stable voxels (stable across the six presentations for all but one concept in the training set) in a cuboid around the centroid of the factor cluster. Because averaging was done over five voxels, clusters smaller than five voxels were excluded from this analysis resulting in a set of 30 clusters.) The model weights were then applied to the ratings of the left-out concept to predict the activation pattern for that concept.

## Comparison between faculty representations acquired in the current study and student representations from a previous study

Comparisons between the two groups were based on only the 15 elementary concepts that had been presented to both groups. First, separate factor analyses of the faculty group and the student group each produced four factors. Each factor was associated with 2 to 5 clusters. The subsequent classification of group membership used a union of the factor cluster locations from these two factor analyses, provided that the cluster was present in at least four participants from the contributing group. This resulted in a set of 11 clusters. The activation level associated with a cluster was the mean activation of the six voxels that were most stable for the 15 concepts (across presentations) from that participant in that cluster.

To identify the most discriminating concepts, a reiterative procedure analogous to stepwise regression was performed. In the first iteration, the group classification was performed using only one concept at a time, determining which single concept of the 30 resulted in the highest classification accuracy. In the second iteration, the classification was performed using pairs of concepts, namely the single concept that produced the highest accuracy in the first iteration as well as each of the 29 other concepts. All pairs that produced at least as high an accuracy as achieved on the previous iteration were explored in the third iteration, where triplets of concepts were used, namely the pairs that produced the highest accuracy in the previous iteration plus each of the remaining 28 concepts. This stepwise addition of discriminating concepts continued until adding any one of the remaining concepts resulted in a decrease in accuracy.

## Reporting summary

Further information on research design is available in the Nature Research Reporting Summary linked to this article.

## DATA AVAILABILITY

All data will be made available from the corresponding author upon reasonable request.

## ACKNOWLEDGEMENTS

We thank the Carnegie Mellon Physics Department faculty for their outstanding cooperation. Additionally, we thank Barry Luokkala for providing corroborating opinions of an early description of the FA dimensions. Barry Luokkala, Marie Amalric, and Vladimir Cherkassky provided comments on earlier drafts of the manuscript. And we thank Vladimir Cherkassky for helpful discussions of the statistical analyses. This research was supported by the Office of Naval Research Grant N00014-16-1-2694 and National Science Foundation Grant BCS-1748897.

## AUTHOR CONTRIBUTIONS

RM, RS, and MJ contributed to the conceptualization, design, data analysis, interpretation, and write-up of the study. RS additionally developed the set of suitable stimulus concepts and the interpretations of the FA dimensions.

## COMPETING INTERESTS

The authors declare no competing interests.

## ADDITIONAL INFORMATION

**Supplementary information** The online version contains supplementary material available at https://doi.org/10.1038/s41539-021-00107-6.

**Correspondence** and requests for materials should be addressed to Robert A. Mason.

**Reprints and permission information** is available at http://www.nature.com/reprints

**Publisher's note** Springer Nature remains neutral with regard to jurisdictional claims in published maps and institutional affiliations.